\pdfoutput=1
\documentclass{raa}
\usepackage{graphicx,times}             %for PS/EPS graphics inclusion, new
\usepackage{natbib}
\usepackage{amssymb,amsmath}
\bibpunct{(}{)}{;}{a}{}{,}
\usepackage[figuresright]{rotating}
\usepackage{tabularx}
\usepackage{color}
\usepackage{enumitem}
\usepackage{adjustbox}
\usepackage[pagebackref=true]{hyperref}
\newcommand{\HI}{H\,{\sc i}}
\usepackage{multirow}
\usepackage{hyperref}
\usepackage{url}
\begin{document}
 
\title{New H\,{\sc i} observations Toward the NGC~5055 Galaxy Group with FAST}
\volnopage{Vol.0 (20xx) No.0, 000--000}      %%preserved for Editor. DOn't remove!
\setcounter{page}{1}          %%starting page, preserved for Editor. DOn't remove!
   \author{Xiao-Lan Liu 
      \inst{1,2,3 }
   \and Ming Zhu
      \inst{1, 2, 3}
   \and Jin-Long Xu
      \inst{1, 2, 3}
   \and Peng Jiang
      \inst{1, 2, 3}
   \and Chuan-Peng Zhang
      \inst{1, 2, 3}  
   \and Nai-Ping Yu
      \inst{1, 2, 3 }  
   \and Jun-Jie Wang
      \inst{1, 2, 3}
     \and Yan-Bin Yang
      \inst{4}    
   }

 \institute{
    National Astronomical Observatories, Chinese Academy of Sciences, Beijing 100101, China; {\it liuxiaolan@bao.ac.cn} \\
\and
    Guizhou Radio Astronomical Observatory, Guizhou University, Guiyang 550000, China;
\and
    CAS Key Laboratory of FAST, National Astronomical Observatories, Chinese Academy of Sciences, Beijing 100101, China; 
\and
    GEPI, Observatoire de Paris, Universite PSL, CNRS, Place Jules Janssen, 92195 Meudon, France. \\
\vs\no
   {\small Received 20xx month day; accepted 20xx month day}}

\abstract{
We report a new high-sensitivity H\,{\sc i} mapping observation of the NGC~5055 galaxy group over an area of $1.^\circ5\times0.^\circ75$ with the Five-hundred-meter Aperture Spherical radio Telescope (FAST). Our observation reveals that the warped H\,{\sc i} disk of NGC~5055 is more extended than what previously observed by WSRT, out to $ 23.'9$ (61.7 kpc). 
The total \HI\ mass of NGC~5055 is determined to be $\rm\sim 1.1\times10^{10}\,M_\odot$.
We identified three H\,{\sc i} clouds with H\,{\sc i} masses of the order of $\rm \sim 10^7\,M_\odot$ at the southeastern edge of the H\,{\sc i} disk, as well as a candidate high-velocity cloud with an H\,{\sc i} mass of $\rm (1.2\pm0.5) \times10^6\,M_\odot$ to the north of NGC~5055.
The HI content of UGCA 337 is robustly detected for the first time by the FAST observations. It has a narrow HI linewidth of $W_{50}=17.4\pm3.8$ km s$^{-1}$ with a total \HI\ mass of ($\rm 3.5\pm0.3)\times10^6\,M_\odot$. Comparing the gas content and g-r color of UGCA~337 with typical low-mass dwarf galaxies, UGCA~337 appears relatively gas-poor despite its blue colour. This suggests that UGCA~337 may have undergone gas stripping in the past. 
We also analyzed the possible origin of the diffuse \HI\ clouds located at the outskirts of NGC~5055, and speculate that they might be the remnant features of a merger event in the past. 
\keywords{galaxies: evolution --- galaxies: structure --- galaxies: individual --- galaxies: kinematics and dynamics --- galaxies: interactions}}

   \authorrunning{Xiao-Lan Liu et al }            %author_head in even pages
   \titlerunning{ H\,{\sc i} observations toward NGC~5055 with FAST}  % title_head in odd pages

   \maketitle

%________________________________________________ sections below
%
\section{Introduction} \label{sec:intro}        %% first-level  sections will be auto-capitalized

Understanding how massive galaxies, such as our Milky Way, accumulate mass to support their almost constant star formation rates over billions of years is crucial to understanding how galaxies evolve. According to the $\Lambda$-Cold Dark Matter paradigm of cosmology, galaxies are expected to form in a hierarchical manner. Gas and dark matter accumulate in dark matter halos, merging and condensing into larger structures, which leads to the formation of stars and the growth of the galaxy. High-sensitivity observations of massive disk galaxies have the potential to provide new insights into the history of their formation and evolution.

NGC 5055, also known as the Sunflower Galaxy M63, is a Milky Way-type SAb(c) galaxy with an optical disk size of about $5.'5$ ($R_{25}$) \citep{Vaucouleurs+etal+1976}. The spiral galaxy NGC~5055 is located at a distance of 8.87 Mpc \citep{McQuinn+etal+2017} and is accompanied by approximately 20 dwarf galaxies, as identified by the deep optical observations \citep[e.g.][]{Karachentsev+etal+2020}. This group is a relatively isolated galaxy group, with no large galaxies in the surrounding $3^\circ \times 3^\circ$ region \citep{Muller+etal+2017, Karachentsev+etal+2020}. Within the group, only a few dwarfs were identified with radial velocities in the range of $\pm 300$ km s$^{-1}$ with respect to that of NGC~5055, among which UGC~8313 and UGCA~337 were determined as the real satellites of NGC~5055 through the measurement of their distances \citep{Carlsten+etal+2022}. However, most of the other faint dwarfs in the group lack velocity or distance information \citep{Karachentsev+etal+2020,Carlsten+etal+2022}, making it difficult to determine their roles in the evolution of NGC~5055.

NGC~5055 has been extensively studied using high-resolution \HI\ line observations at 21-cm carried out by the Westerbork Synthesis Radio Telescope (WSRT) \citep{Battaglia+etal+2006, Kamphuis+etal+2022}. These observations revealed a significantly more extended \HI\ disk with pronounced kinematic warps beyond the inner optical disk. Deep optical observations have found very faint stellar loops across NGC~5055, some of which had no \HI\ associations detected \citep{Chonis+etal+2011,Karachentsev+etal+2020}. 
%\yb{please add something here, about the studies of the deep optical observations, the stellar loops, and their comparison with HI observation (if possible)}
%\citet{Battaglia+etal+2005} discovered a declining rotation curve that drops at the end of the optical disk. 
\citet{Kamphuis+etal+2022} reported three \HI\ sources on the outskirts of NGC~5055 through analyzing the deep WSRT Hydrogen Accretion in LOcal GAlaxieS (HALOGAS) Survey \footnote{\url{https://www.astron.nl/halogas/data.php}} \citep{Heald+etal+2011}, including a possible cloud candidate located to the northeast of NGC~5055 and two \HI\ counterparts of the nearby dwarf galaxies UGC~8313 and UGC~8365.

Interferometer arrays have the advantage of resolving small structures of galaxies in principle, but can suffer from the so-called "negative bowl" artifacts indicating the loss of short-spacing information \citep{Wang+etal+2023}. In contrast, single-dish radio telescopes can compensate for this deficiency. In recent years, radio telescopes with large collecting areas, such as the Five-hundred-meter Aperture Spherical radio Telescope (FAST), have detected faint gas in or around several large spiral galaxies \citep[e.g.,][]{Zhu+etal+2021,Xu+etal+2021,Wang+etal+2023,Yu+etal+2023}. These findings have cast new light on the formation and evolution of galaxies.  In this paper, new H I observation have been carried out towards the NGC 5055 galaxy group using FAST. We describes the observation and data reduction in Section 2. The results and the kinematic analysis of NGC~5055 are presented in Section 3. Section 4 discusses the possible origin of the observed \HI\ around NGC~5055. While Section 5 provides a summary of the main results in this paper.

\section{Observation and Data reduction} \label{sect:Obs}
We carried out a high-sensitivity \HI\ observation of NGC~5055 and its surroundings over an area of $1.^\circ5\times0.^\circ75$ in 2022 July, as part of the FAST All Sky \HI\ survey \citep[FASHI,][]{Zhang2024SCPMA..6719511Z}. We adopted the multi-beam on-the-fly (MultiBeamOTF) observation mode with the 19-beam array receiver which provides a half-power beam width (HPBW) of $2.'9$ at 1.4 GHz per beam \citep{Nan+etal+2011,Jiang+etal+2019,Jiang+etal+2020}. The on-sky scanning speed was set to $15''$ per second with an integration time of 1 s. Spectra were observed in the dual polarization mode of the front end, and digitized by the backend Spec(W) in 65536 channels over a bandwidth of 500 MHz. The latter provides a frequency resolution of 7.629 kHz at 1.4 GHz, corresponding to a velocity resolution of 1.611 km s$^{-1}$. During the observation, the mean system temperature was 20 K.  The pointing accuracy of FAST is better than $8''$. We observed NGC~5055 twice with a total observation time 1.28 hr. For calibration, a standard noise signal, $T_{\rm cal}$ (typically 10~K), was regularly injected into the signal path every 32 s with a duration of 1 s. The raw data was reduced using the Python-based pipeline HIFAST \citep{Jing2024arXiv240117364J}. The reduced data cube has a pixel scale of $1'\times 1'$ and a channel resolution of 1.611 km~s$^{-1}$ in spectrum. The final data-cube has a mean noise about 1.04 mJy beam$^{-1}$ per channel, corresponding to a 3$\sigma$ \HI\ column density of $\rm 5.2\times10^{17}\,cm^{-2}$ in a line width of 20 km s$^{-1}$ \citep{Wang+etal+2023}.

In addition, we utilized SOFIA \citep{Serra+etal+2015MNRAS.448.1922S} to create a detection mask for the \HI\ emission in the final data cube, which was applied during the analysis of the \HI\ moment maps. The smooth+clip source-finding algorithm was implemented by applying a 5$\sigma$ threshold and smoothing kernels of 0, 3, and 5 pixels in both space and velocity dimensions. The final detection mask was generated by ensuring a detection reliability of 0.99.

\section{Results} \label{sec:results}
\subsection{HI map of the NGC~5055 galaxy group} \label{subsec:satellites}
\begin{figure*}[htb!]
   \centering
   \hspace{-8mm}
  \includegraphics[angle=0,width=0.75\columnwidth]{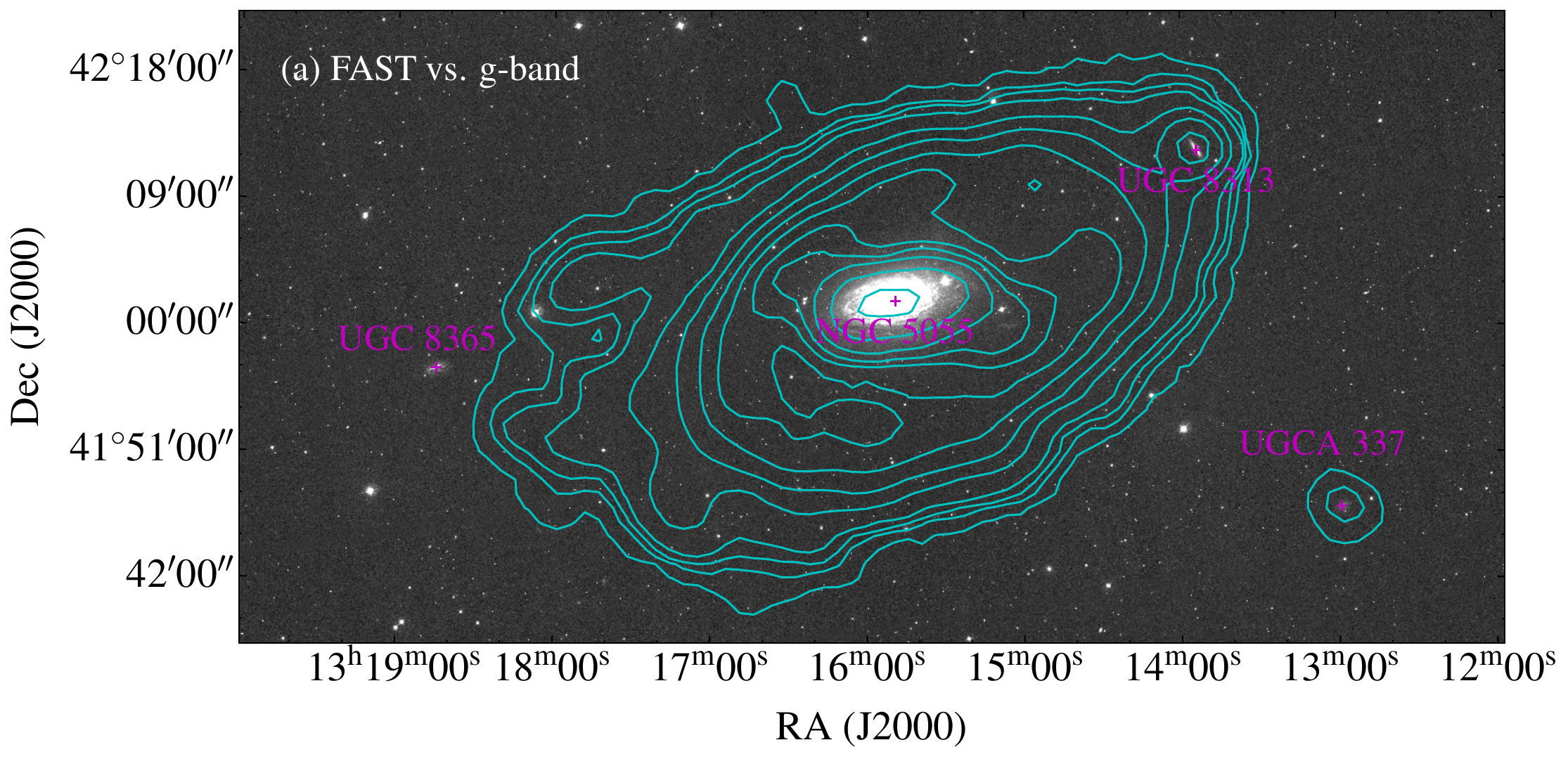}
  \includegraphics[angle=0,width=0.8\columnwidth]{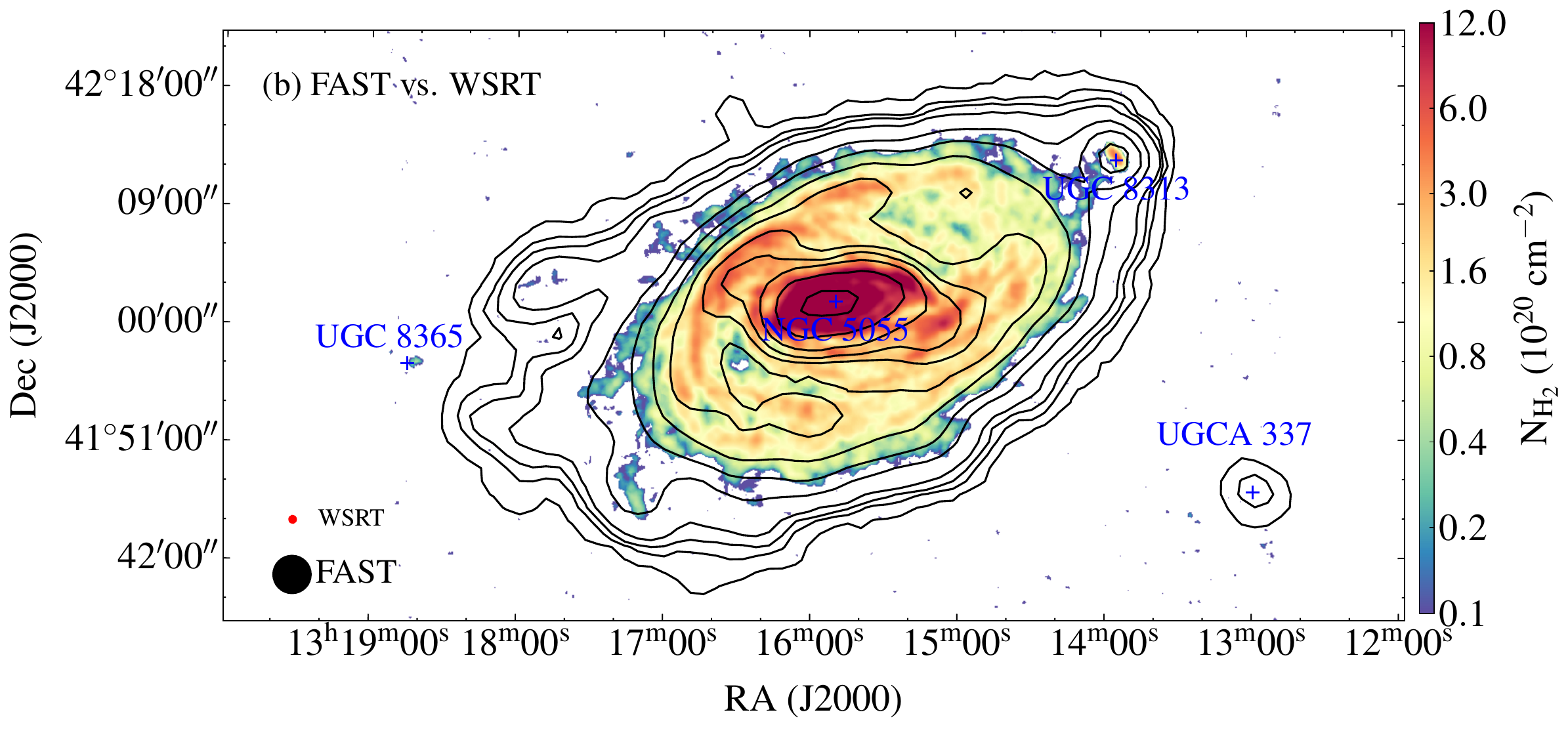}
   \caption{H\,{\sc i} column density contours in the velocity range of 276.0-718.8 km s$^{-1}$ superimposed on the SDSS g-band optical image of the NGC~5055 galaxy group (a) and the HALOGAS survey data (b). The contours are 0.1 ($3\sigma$), 0.3, 0.6, 1.0, 3.3, 6.8, 12.8, 20.6, 31.4, 41.2, 55.5, 82.4, 123.4, 156.0$\times10^{19}\,\rm cm^{-2}$ respectively. The plus symbols indicate the optical centers of galaxies NGC~5055, UGC~8313, UGCA~337 and UGC~8365, respectively. The beamsizes of FAST and WSRT are shown in the bottom-left corner of panel (b). }
   \label{DeepOptical}
   \end{figure*}

Figure~\ref{DeepOptical}a presents H\,{\sc i} column density map in contours of the NGC~5055 galaxy group, which is overlaid on the SDSS g-band optical image. The column density is derived from its moment-0 map integrated over the velocity range of $276.0-718.8$ km s$^{-1}$. There are three dwarf galaxies located adjacent to NGC 5055. UGC~8313 and UGCA~337 had been recognized as the satellites of NGC~5055 \citep{Carlsten+etal+2022} at projected distances of $\rm \sim 61.7\, kpc$ to the northwest of NGC~5055 and $\rm \sim 90\, kpc$ to the southwest respectively. UGC~8365 is located on the east of NGC 5055, and has a system velocity of 1260 km s$^{-1}$\citep{Driel+etal+2016, Kamphuis+etal+2022}. We estimated the distance of UGC~8365 to be $20.8\pm2.1\,\rm Mpc$ using the Cosmicflows-3 method \citep{Kourkchi+etal+2020}, which is not associated with that of NGC 5055. Thus, we suggest that UGC~8365 is a background galaxy of NGC 5055.

%\subsubsection{Comparison with HALOGAS survey by WSRT}
Figure~\ref{DeepOptical}b shows an \HI\ column density overlay map of NGC~5055, obtained from the FAST observation data and the HALOGAS survey data. Compared to the HALOGAS survey data, we found that the FAST observations reveal more extended \HI\ structures around NGC~5055. It makes the observed \HI\ disk of NGC~5055  out to $ 23.'9$ (i.e.,  $\rm 61.7\,kpc$), which is corrected for resolution utilizing the expression of $r_{\rm eff} = \sqrt{\rm R_{HI}^2-R_B^2}$, where $\rm R_{HI}$ is the length of the semi-major axis measured from the \HI\ column density map in the \HI\ disk over $3\sigma$ and $\rm R_{B}$ is the half of the HPBW of FAST. This \HI\ disk measured by FAST is $\sim 25\%$ larger than that of the full WSRT data observed by \citet{Battaglia+etal+2006}. 

\begin{figure}[h!]
   \centering
  \includegraphics[angle=0,width=0.5\columnwidth]{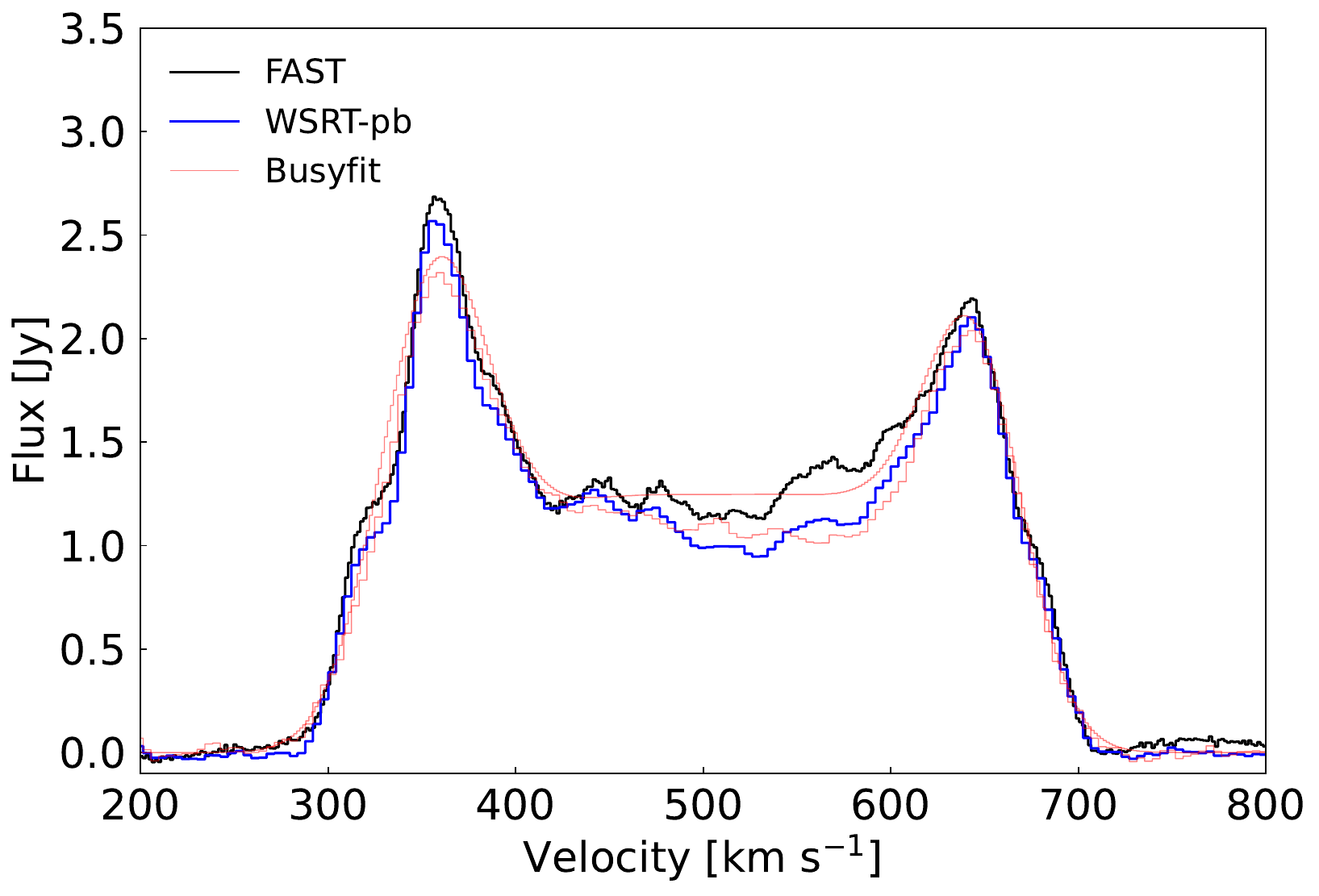}
   \caption{ Comparison of the integrated H\,{\sc i} profile over NGC~5055 from the FAST map (black line) and the pb-corrected HALOGAS map (blue line). The red lines represent the fittings from the busy function code \citep{Westmeier+etal+2014}. }
   \label{intflux5055}
   \end{figure}

Figure~\ref{intflux5055} compared the total \HI\ flux of NGC~5055 from FAST observation (black line) and the primary beam(pb)-corrected HALOGAS survey (blue line). We find the pb-corrected HALOGAS survey revealed much less HI flux, by $22.4\%\pm0.6\%$ percent, compared to that observed by FAST (see Table~1). Taking into account of the $\sim 10\%$ systematic difference between the two telescopes and the $\sim 10\%$ flux calibration uncertainty \citep{Wang+etal+2023}, our observed \HI\ flux for NGC~5055 is still in excess of the HALOGAS survey. This is because HALOGAS survey was performed with the interferometer array WSRT, which could suffer from the zero-spacing problem and miss some fluxes from the extended gas component. This effect has been studied in details in the cases of NGC~4631 \citep{Wang+etal+2023}. 
\begin{table*}[h!]
\bc
\begin{minipage}[]{120mm}
\caption{The observed parameters for \HI\ sources in the NGC~5055 galaxy group.}\end{minipage}
\setlength{\tabcolsep}{5pt}
\small
 \begin{tabular}{ccccccccccccc}
  \hline\noalign{\smallskip}
source  & $\rm log(M_{\rm H\,{\sc I}})$  & $ S_{\rm H\,{\sc I}}$       &$S_{\rm peak}$    &$ \upsilon_{\rm sys}$  & $W_{50}$  & $W_{20}$ &Velocity range    \\%&${r_{\rm eff}}^*$
             &$ \rm(M_\odot)$ & $\rm(mJy \,km\, s^{-1})$ & $\rm (mJy)$  & $\rm(km \,s^{-1})$      & $\rm(km\,s^{-1})$   & $\rm(km\, s^{-1})$   & $\rm(km \,s^{-1})$  \\%&$\rm [arcmin]$ 
(1)       & (2)      & (3) & (4) & (5)  & (6)     & (7)  & (8) \\
  \hline\noalign{\smallskip}
\multicolumn{8}{c}{FAST} \\ \hline

NGC~5055        & $10.03(0.01)$ &$574.8(17.3)\times10^3$     &2390.0(131.2)     &492.8(0.4) &346.7(0.5) &383.5(0.5)  &276.0-718.8 \\
UGCA~337        & 6.55(0.04)    &194.7(17.5)     &10.8(1.7)       &550.0(1.0)   &17.4(3.8)  &25.3(5.6)   &533.7-557.8 \\
c1              & 7.43(0.01)    &1449.0(43.2)    &35.3(1.8)       &564.5(0.8)   &39.3(1.9)  &58.2(2.5)   &536.9-601.3\\
c2              & 7.36(0.01)    &1239.0(36.2)    &33.0(1.3)       &621.6(0.6)   &37.0(1.4)  &51.1(1.9)   &583.6-651.2 \\
c3              & 7.09(0.02)    &661.6(35.0)     &27.2(2.5)       &616.5(0.9)   &23.6(2.9)  &34.0(4.4)   &590.0-657.7  \\
H1              & 6.09(0.17)    &66.0(25.5)      &2.9(1.1)        &322.8(3.7)   &21.5(11.3) &33.0(17.0)   &316.3-329.2 \\
\hline
\multicolumn{8}{c}{HALOGAS (after bp correction)} \\ \hline
NGC~5055        & 9.94(0.02)    &$469.6(19.8)\times10^3$  &1727.0(134.1)   &497.9(0.6) &354.2(2.7) &389.1(1.9)  &276.0-718.8  \\                             
\noalign{\smallskip}\hline
\end{tabular}
\ec
\label{tab1}
\end{table*}

\begin{figure}[htbp]
  \begin{minipage}[b]{0.5\linewidth}
   \centering
  \includegraphics[width=1.0\columnwidth]{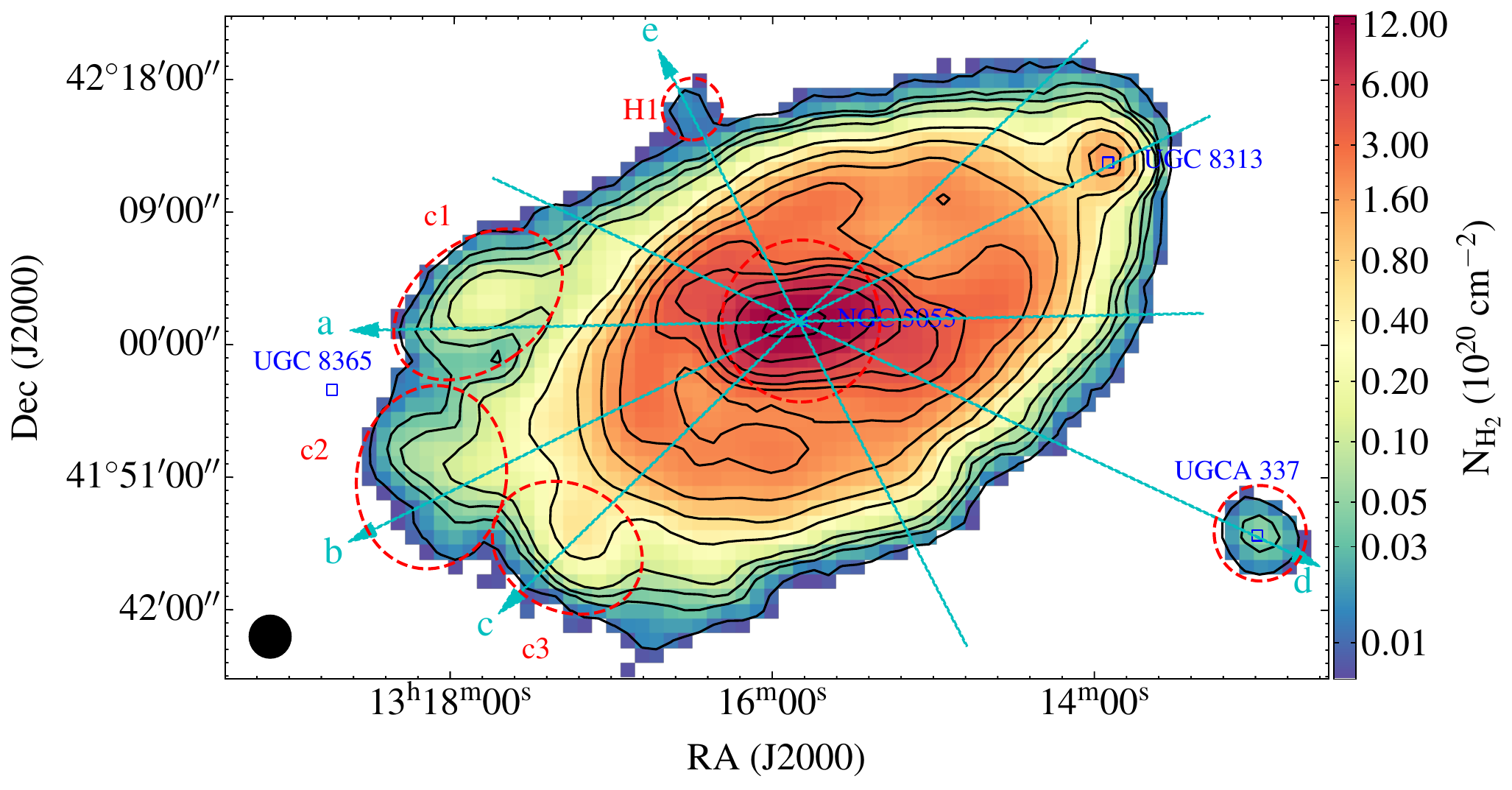}
  \end{minipage}
  %\hspace{-3mm}
  \begin{minipage}[b]{0.5\linewidth}
  \centering
  \includegraphics[width=0.9\columnwidth]{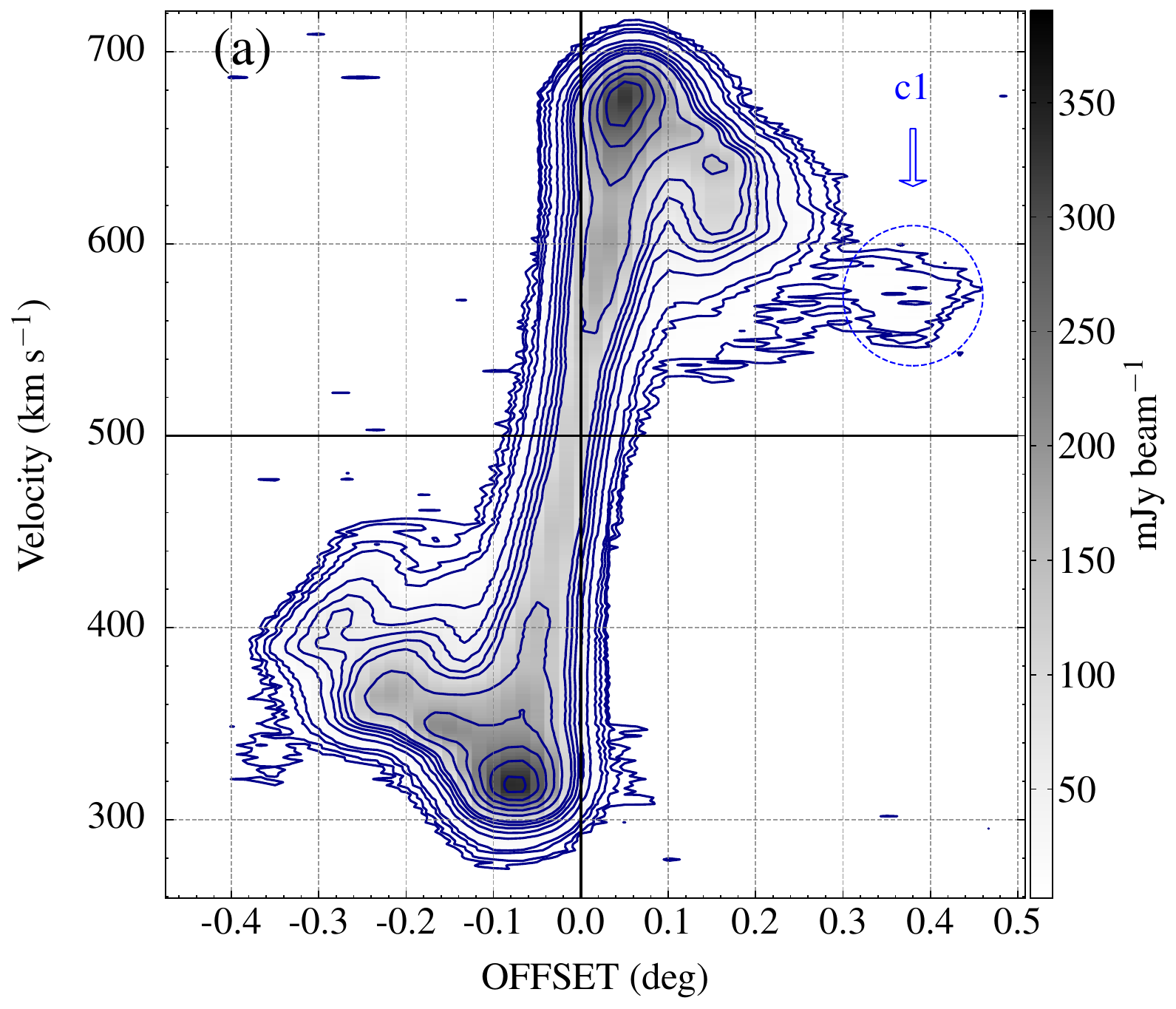}
  \end{minipage}
  \\
  \begin{minipage}[b]{0.5\linewidth}
  \centering
  \includegraphics[width=0.9\columnwidth]{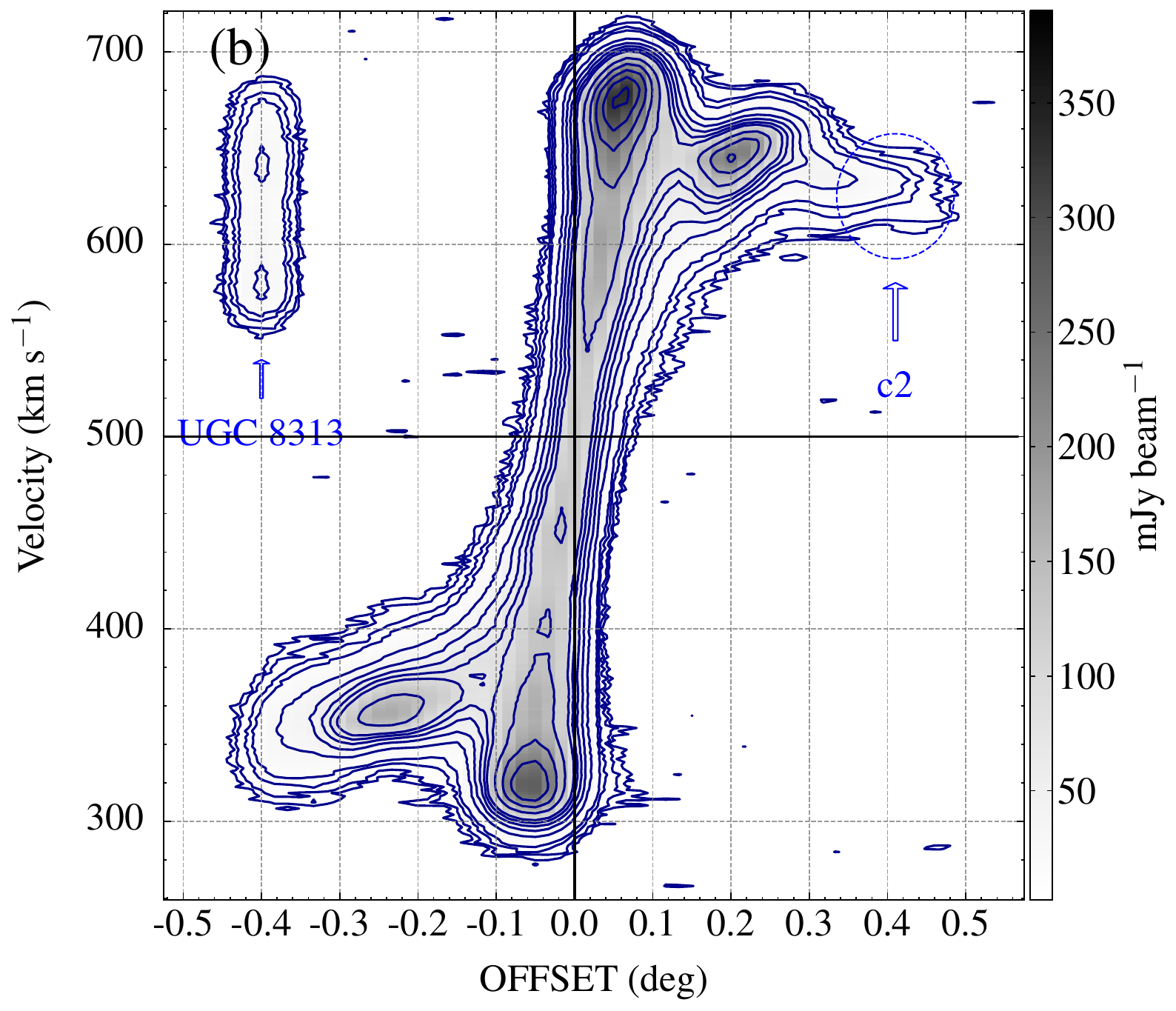}
  \end{minipage}
  %\hspace{-3mm}
  \begin{minipage}[b]{0.5\linewidth}
  \centering
  \includegraphics[width=0.9\columnwidth]{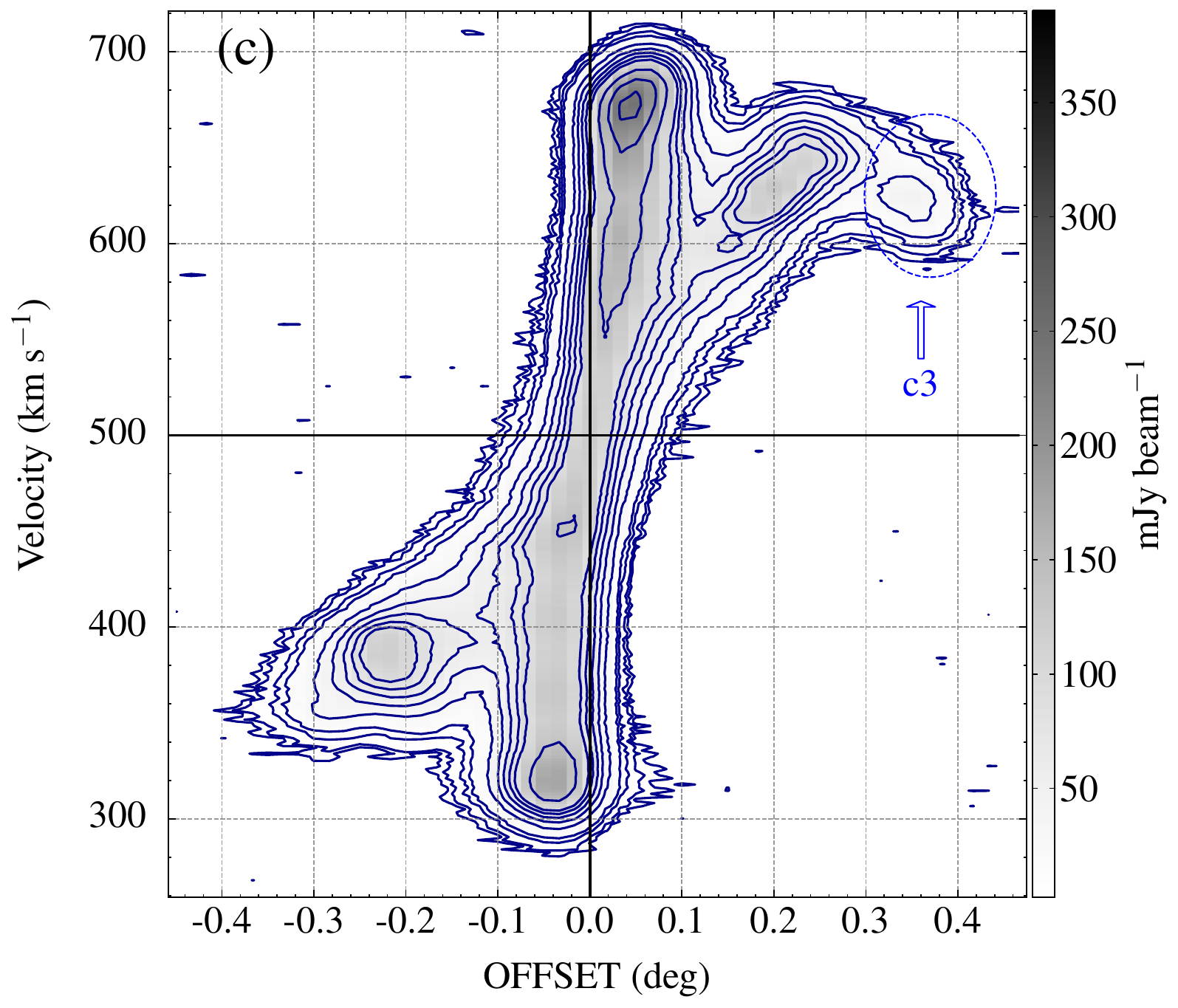}
  \end{minipage}
  \\
  \begin{minipage}[b]{0.5\linewidth}
  \centering
  \includegraphics[width=0.9\columnwidth]{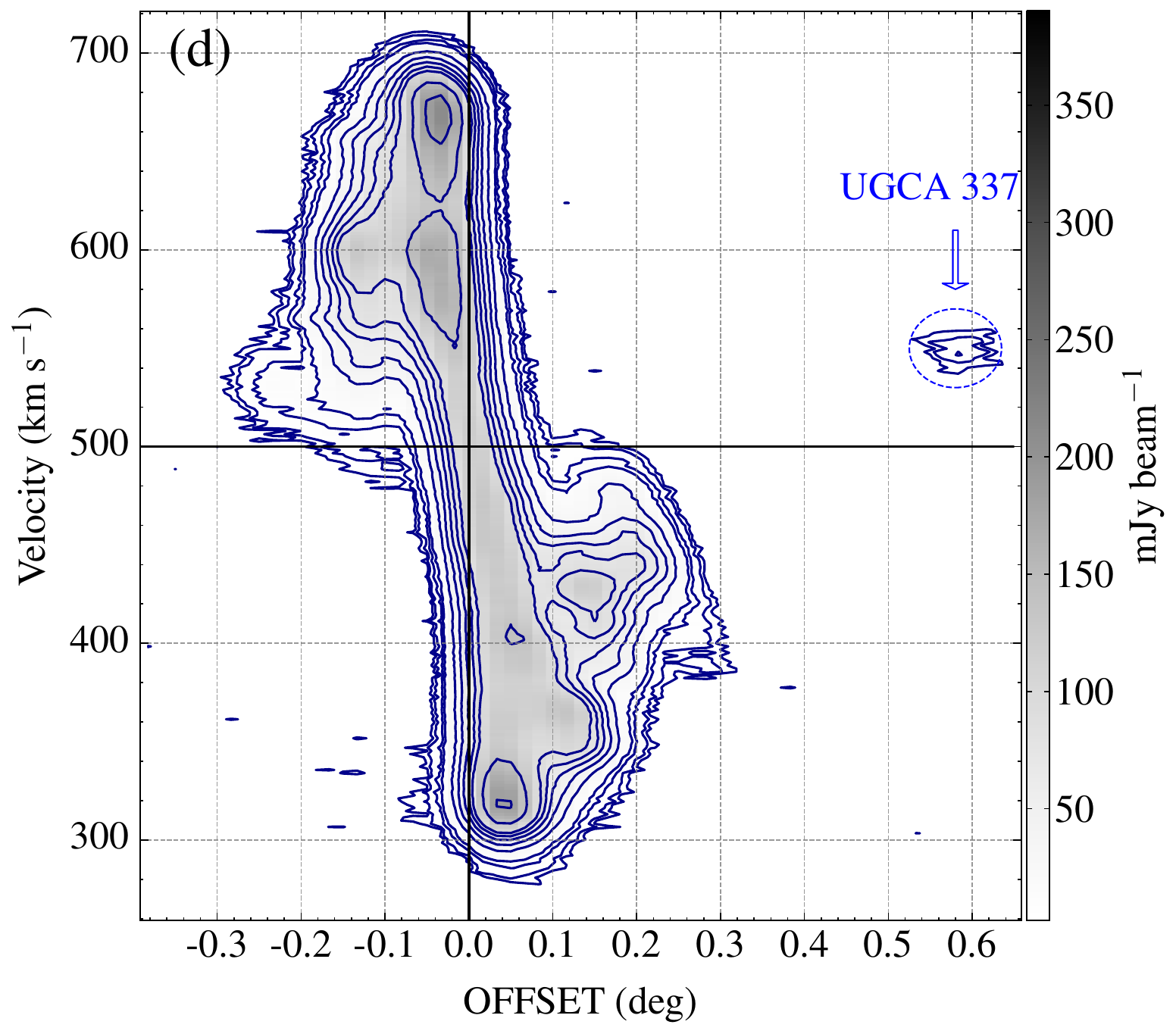}
  %\hspace{0.5mm}
  \end{minipage}
 % \hspace{-3mm}
  \begin{minipage}[b]{0.5\linewidth}
  \centering
  \includegraphics[width=0.95\columnwidth]{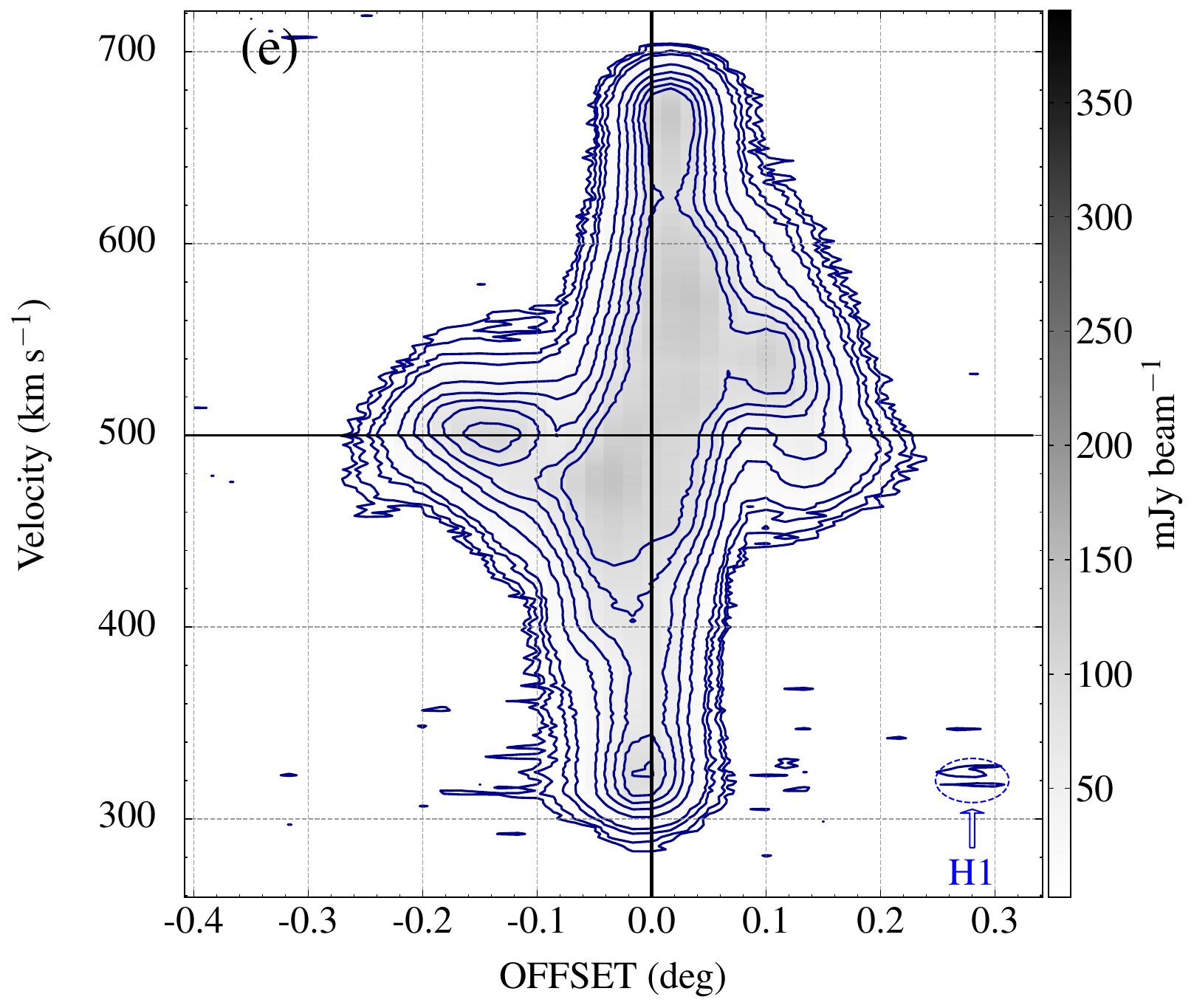}
  \end{minipage}
\caption{Five PV plots along different directions. The top-left frame shows H\,{\sc i} column density maps (black contours+color map) of the NGC~5055 galaxy group. The \HI\ contour levels and symbols are the same as those in Figure~\ref{DeepOptical} respectively. The five cyan dashed arrows indicate the orientation along which to make the PV diagrams in panels (a), (b), (c), (d), (e), respectively. The dashed red ellipses and circles in each panel represent the possible \HI\ substructures c1, 2, 3, H1, as well as the possible \HI\ component of UGCA~337 respectively. The blue contours in panels (a)-(e) are $3\sigma$, $5\sigma$, $10\sigma$, $15\sigma$, $35\sigma$, $55\sigma$, $75\sigma$, $95\sigma$, $115\sigma$, $175\sigma$, $235\sigma$, $295\sigma$, $355\sigma$, $415\sigma$, $475\sigma$, $535\sigma$, $595\sigma$, $655\sigma$, $715\sigma$, respectively. The mean $1\sigma$ value is equal to 0.78 mJy beam$^{-1}$ for the slices taken within the $3^\prime$ width centered on the paths a-e.}
   \label{pv}
\end{figure}

\begin{figure}[t!]
   \centering
  \includegraphics[scale=0.3]{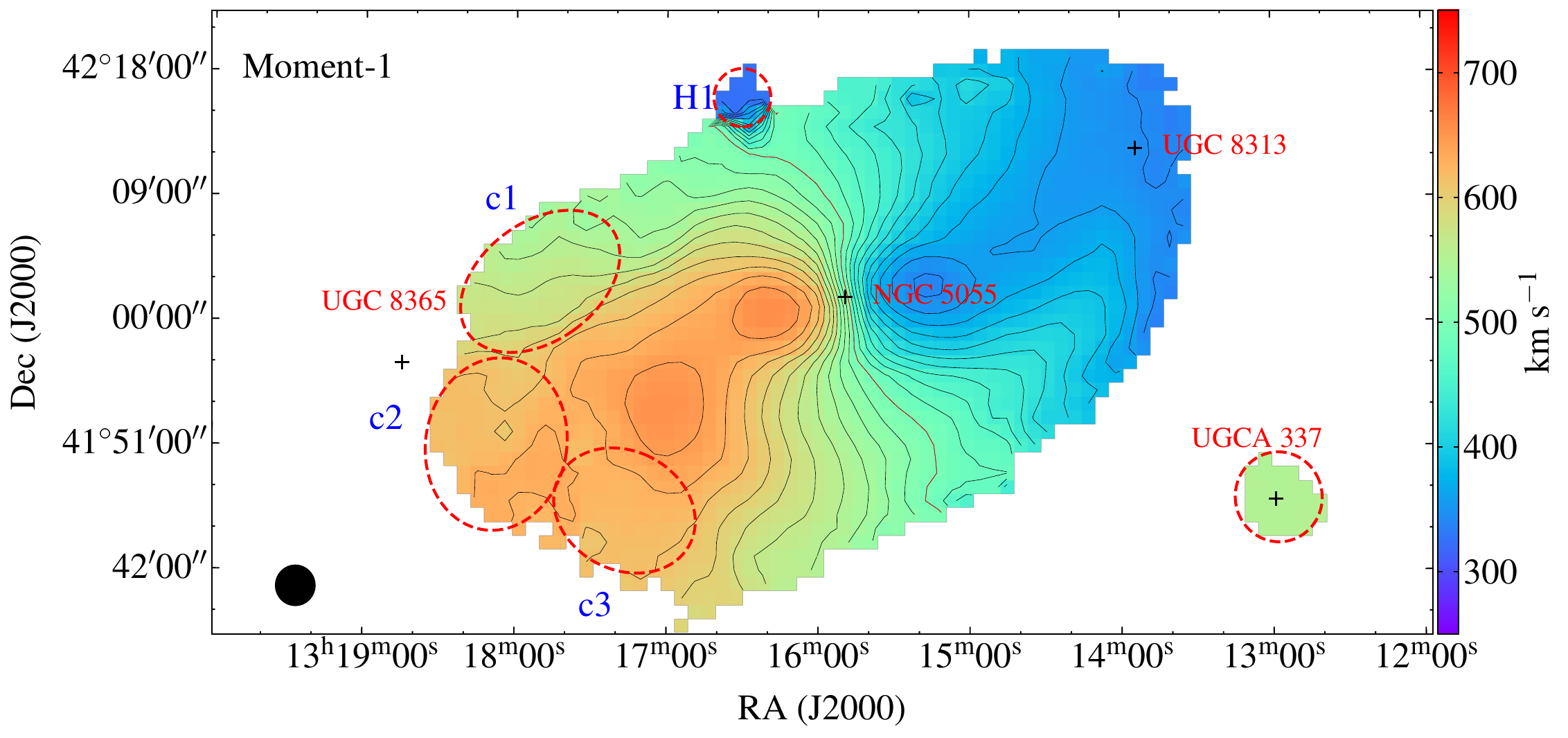}
   \caption{H\,{\sc i} intensity-weighted velocity field (i.e., the first moment map) of NGC 5055. UGC~8313 has been excluded in this calculation. Velocity contours start from 297.8 to 672.8 km s$^{-1}$ in steps of 15 km s$^{-1}$. The red solid line denotes the system velocity of NGC~5055. Dashed ellipses/circles and pluses represent the same features as those in Figure~\ref{pv}. The beamsize of FAST is displayed in the bottom-left corner. }
   \label{velfield}  
\end{figure}

\subsection{HI kinematics of NGC 5055}
In Figure~\ref{pv}, we explore the position-velocity (PV) slices along five directions, i.e., the cyan dashed arrows from $a$ to $e$ (the same names apply for the panels). 
The slice thickness is taken as 3 spaxels, i.e., 3.'0, corresponding to the beamsize. 
In each PV, the position axis is chosen to be centered on the center of NGC 5055, and increasing along the arrow direction. 
These slices were selected to investigate the kinematic behaviors of dwarf galaxies and some possible substructures (i.e., c1, 2, 3, and H1) in the \HI\ map around the main (inside a elliptical radius of 0.$^\circ$3) \HI\ disk of NGC 5055.
Panel (a) shows a slice of PV passing a possible substructure of c1, which seems to be a detached substructure with a noticeable smaller velocity compared to that of the main \HI\ disk emissions. In the positive part of the PV, the c1 shows some broad connections in space to the main \HI\ disk over a wider velocity range. 
One may also notice the counterpart (i.e., at the same symmetric distance) of the c1 in the negative part of the PV, which appears not detached from the main \HI\ disk, but with strong lagging in velocity (over about 50 km s$^{-1}$). Although these spatial and kinematic features of these lagging \HI\ emissions and c1 could result from the beam smearing effect at high inclination, it is important to note that the ridge line of the disk flux in panel (a) points to c1, indicating that they are likely to be tidal gas around NGC~5055.  %according to the study by \citet{Marasco+etal+2019}.mostly lagging with respect to that of the main \HI\ disk. 
In the panel (b), UGC 8313 is seen at a counter-rotating velocity with respect to the disk of NGC 5055, and well detached. The possible substructure c2 seems more connected to the main \HI\ disk, and its counterpart appears well as a continuity of the main \HI\ disk.
Panel (c) shows that the possible substructure c3 has similar properties as c2.
Panel (d) shows the clear detection of \HI\ gas associated with the low surface brightness (LSB) galaxy UGCA~337.
In the panel (e), we see that the substructure H1 appears not only as an extra structure from the main \HI\ disk, but also separated in velocity by $\approx \rm -170\,km\,s^{-1}$ to the systemic velocity of NGC 5055. This suggests that it is either a satellite or a high velocity cloud (HVC) of NGC~5055 \citep{Westmeier+etal+2008, Hess+etal+2009,Chynoweth+etal+2011AJ....141....9C}. 

Figure~\ref{velfield} presents the velocity field of NGC~5055 by excluding UGC~8313, as it is superimposed in position with the disk but at a different velocity. Similar to the PV slices in Figure~\ref{pv}, the main \HI\ disk of NGC 5055 presents a regular warped velocity field. At the outskirts of NGC~5055, c1, 2, and 3 exhibit the velocity continuity with the main \HI\ disk of NGC~5055 on the whole, however, there appears to be a slightly perturbed velocity field at the junctions between them. In addition, a velocity gradient is presented between the isolated H1 and the northern edge of the \HI\ disk of NGC~5055, possibly due to the projection effect.

\begin{figure}[htb!]
\centering
\includegraphics[angle=0,width=0.98\columnwidth]{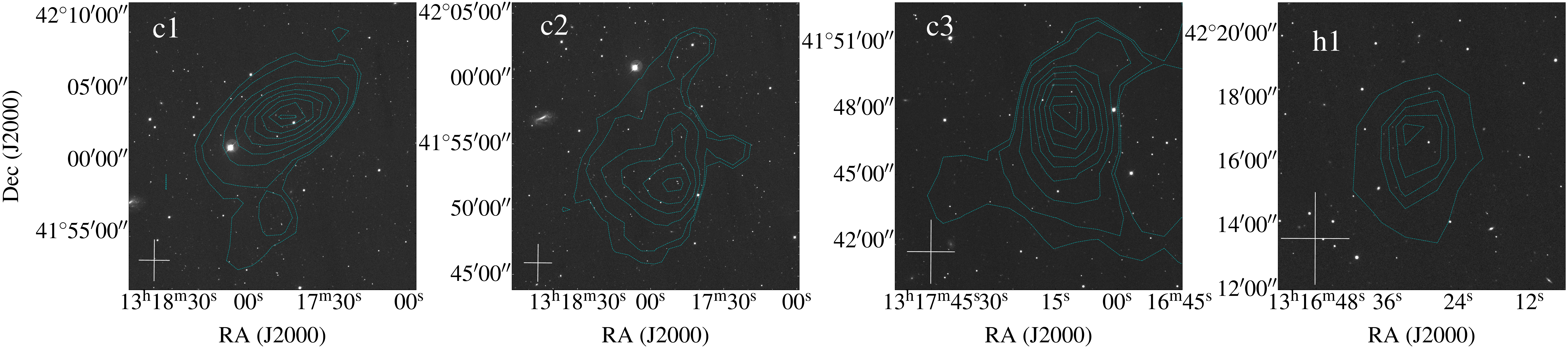}
  %\end{minipage}
  \\
  \vspace{5mm}
  %\begin{minipage}[b]{1.0\linewidth}
  %\centering
  \includegraphics[angle=0,width=0.98\columnwidth]{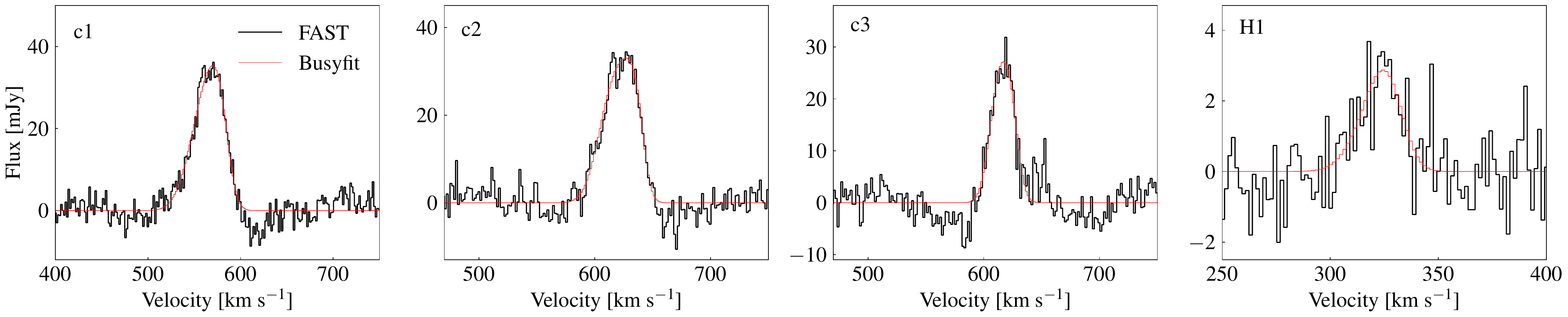}
  %\end{minipage}
  \\
\caption{Top: H\,{\sc i} column density contours superimposed on the SDSS g-band optical image for clouds c1, 2, 3, along with the candidate HVC H1. Black contour levels for cloud c1 are 1.2 ($3\sigma$), 2.0, 4.5, 7.1, 9.7, 12.3, 14.8, 17.4, 20.0$\rm \times10^{18}\, cm^{-2}$, for cloud c2 are 1.2 ($3\sigma$), 2.0, 4.6, 7.1, 9.7, 12.3$\rm \times10^{18}\, cm^{-2}$, for cloud c3 are 2.0 ($3\sigma$), 5.0, 8.0, 10.9, 13.9, 16.9, 19.9, 22.0$\rm \times10^{18}\, cm^{-2}$, and for cloud H1 are 0.6 ($3\sigma$), 0.9, 1.0, 1.2, 1.3$\rm \times10^{18}\, cm^{-2}$. The beam size of FAST is displayed in the bottom-left corner in each panel. Bottom: The corresponding integrated \HI\ spectra lines. The red lines represent the fittings from the busy function code \citep{Westmeier+etal+2014}.}
\label{modelclouds}
\end{figure}

\subsection{Integrated HI of HI clouds}   
%\textcolor{red}{To extracting the \HI\ emission in c1, 2, 3 from the host galaxy NGC 5055, we created a 3D model for NGC~5055 using the 3D tilted ring fitting by the software TiRiFiC \citep{Jozsa2007} with two half disks. Figure~\ref{modeln5055} present the comparisons between the FAST data and the 3D model for NGC~5055. Overall, the model almost reproduces the main gaseous disk of NGC~5055, but does not recover the three \HI\ substructures. This suggests that they may be extraplanar gas with respect to the disk gas in NGC 5055.} \textcolor{red}{after subtracting the TiRiFiC model for NGC 5055}
Using the associated velocity ranges from the PV diagrams (see Table 1), we created \HI\ column density maps and integrated fluxes of the \HI\ substructures c1, 2, 3, and H1, which are presented in Figure~\ref{modelclouds}. The \HI\ column density maps over 3$\sigma$ were superimposed on the SDSS g-band optical images to identify the optical counterparts. In Figure~\ref{modelclouds}, each of the four \HI\ substructures exhibits a coherent structure with a Gaussian component, and appears to lack any visible optical counterparts. Therefore, we can conclude that the \HI\ substructures c1, 2, 3 and H1 are probably coherent gas clouds \citep{Chynoweth+etal+20112011AJ....142..137C} located on the outskirts of the extended gaseous disk of NGC~5055. We compute their total H\,{\sc i} mass through the following equation :
\begin{equation}
 M_{\rm H\,{\sc I}} = 2.355\times10^5 D^2 \int {S_\nu d \upsilon},
 \label{eq1}
 \end{equation}
where $D$ is the distance in Mpc and $\int {S_\nu \rm d \upsilon}$ is the integrated H\,{\sc i} flux in Jy km s$^{-1}$. Assuming the same distance as the host galaxy NGC~5055, clouds c1, 2, 3 and H1 are determined to have a total \HI\ mass of $\rm (2.7\pm0.06)\times10^7\,M_\odot$, $\rm (2.3\pm0.05)\times10^7\,M_\odot$, $\rm (1.2\pm0.06)\times10^7\,M_\odot$, and $\rm (1.2\pm0.5)\times10^6\,M_\odot$ respectively, as shown in column 2 of Table~1. Their total \HI\ mass contributes to less than one percent of that of NGC~5055, which is about $\rm 1.1\times10^{10}\, M_\odot$ as derived from the integrated \HI\ flux presented in Table~1.

\subsection{Integrated HI of the LSB darf galaxy UGCA~337}
\begin{figure}[!h]
   \begin{minipage}[b]{0.45\linewidth}
  \centering
  \includegraphics[angle=0,width=0.95\columnwidth]{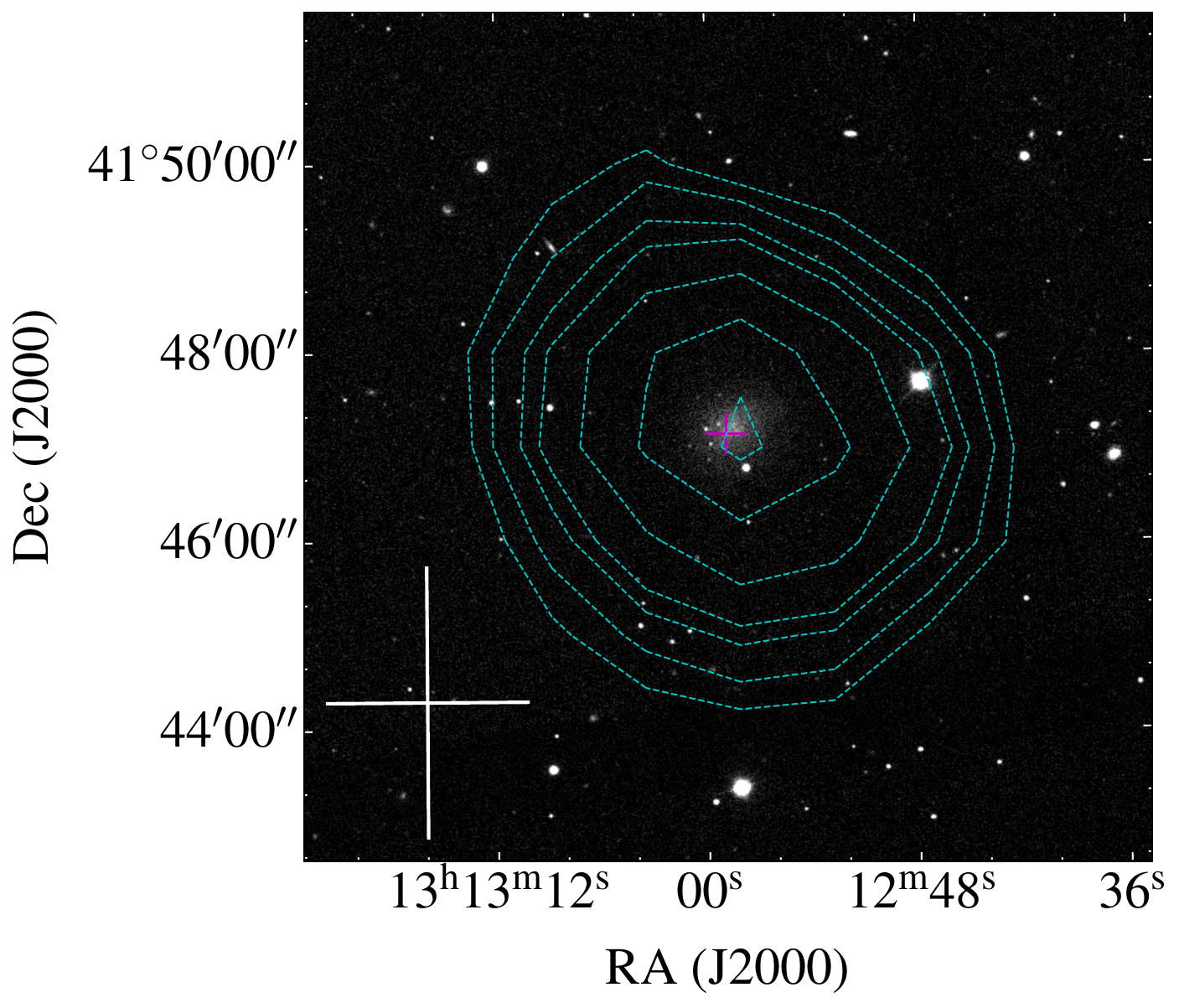}
  \end{minipage}
\begin{minipage}[b]{0.45\linewidth}
  \centering
  \includegraphics[angle=0,width=0.87\columnwidth]{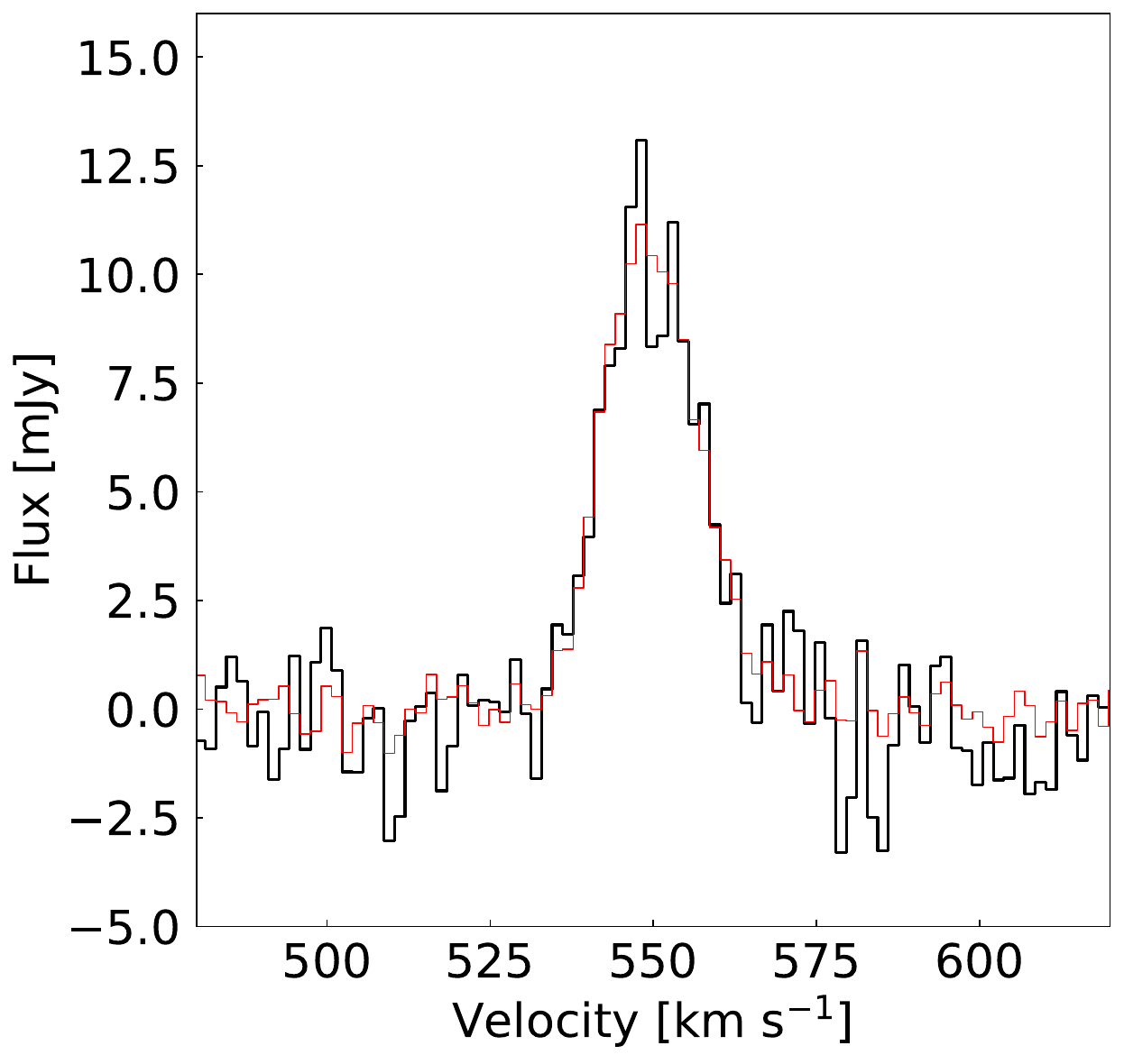}
  \end{minipage}
\caption{Left: H\,{\sc i} column density contours in the velocity range of 533.7-557.8 km s$^{-1}$ superimposed on the SDSS g-band optical image for UGCA~337. The cyan contour levels are 1.0 ($3\sigma$), 1.4, 1.7, 2.4, 3.5 and 4.5$\times10^{18}\rm\,cm^{-2}$. The magenta plus marks the optical center of the galaxy. The beamsize of FAST is displayed in the bottom-left corner. Right: The corresponding integrated \HI\ spectrum for UGCA~337. The red line represents the fitting from the busy function code \citep{Westmeier+etal+2014}.}
   \label{dwarfgalaxies}
   \end{figure}

Figure~\ref{dwarfgalaxies} shows the \HI\ column density contours overlaid on the SDSS g-band optical image and integrated \HI\ spectrum (including the fitting from the busy function code) for UGCA~337. From the fitting results obtained from the busy function code of the integrated \HI\ spectrum in Table~1, FAST detected a narrow single-Gaussian \HI\ component with a system velocity of $\rm 550.0\pm1.0\, km\,s^{-1}$ and a linewidth of $W_{50}\rm \sim 17.4\pm3.8\, km\,s^{-1}$. It is in agreement with the optically heliocentric radial velocity of $\rm 529\pm40\, km\,s^{-1}$ \citep{Alam+etal+2015} and therefore can be identified as the counterpart of UGCA~337. This \HI\ component is missing from the HALOGAS survey. However, in the previous studies for UGCA~337 \citet{Huchtmeier+etal+2009} reported a double-horn \HI\ counterpart with a system velocity of $\rm 460.0\pm5.0\, km\,s^{-1}$ and a linewidth of $W_{50}=234$ km s$^{-1}$, observed by the 100-m radio telescope Effelsberg with a spatial resolution of $9^\prime$. 
Given the facts that UGCA~337 is comparable to the beamsize of FAST but much smaller than that of Effelsberg and the \HI\ component detected by FAST exhibits a higher signal-to-noise ratio, we believe that the \HI\ results of UGCA 337 from the FAST observation are more reliable, which means we got a clear HI detection of this galaxy for the first time. As for the \HI\ detection performed by \citet{Huchtmeier+etal+2009}, it might have been affected by the \HI\ emission from NGC 5055 that has leaked in through an Effelsberg sidelobe \citep{Reich+etal+1978A&A....69..165R}, and thus give the impression of a double-horned profile.

%A comparison with an untargeted or similar population of large galaxy samples will give us comprehensive understanding of the gas nature in UGCA~337. We plot the ratio of $ M_{\rm H\,{\sc I}}/M_\ast$ to $M_\ast$ by comparing it to the complete ALFALFA-SDSS galaxy catalogue \citep{Durbala+etal+2020AJ....160..271D} and nearby LSB galaxies \citep{Honey+etal+2018} in Figure~\ref{relocategalaxies}. 
Cold gas ($M_{\rm gas}/M_{\star}>1$) typically dominates the baryon content in low-mass dwarfs \citep{Lelli2022NatAs...6...35L}. We have determined the total \HI\ mass of UGCA~337 to be $\rm (3.5\pm0.3)\times10^6\,M_\odot$ using equation~\ref{eq1} and the integrated \HI\ flux of $194.7\pm17.5$ mJy km s$^{-1}$. The total gas mass, denoted as $M_{\rm gas}$, can be calculated as $\rm (4.7\pm0.4)\times10^6\, M_\odot$, after accounting for helium's contribution, by multiplying the above estimated value by a factor of 1.33.
The stellar mass ($\rm M_{\star}$) of UGCA~337 is determined to be $\rm (1.7\pm0.5) \times10^8\,M_\odot$ after correcting for distance to 8.87 Mpc \citep{Carlsten+etal+2022}. The obtained $\rm M_{gas}/M_{\star}$ in UGCA~337 is about $0.03\pm0.008$, implying that this dwarf galaxy is extremely gas-poor. Additionally, the dwarf galaxy exhibits a g-r color of  $0.61\pm0.13$ \citep{Carlsten+etal+2022} and a star formation rate of $4.8\times10^{-5}\,\rm M_\odot\, yr^{-1}$ as estimated from the H$\alpha$ observation \footnote{http://www.sao.ru/lv/lvgdb/object.php?name=UGCA337\&id=696}, indicating a rather low level of star formation despite its blue color. Such findings differ from the usually low-mass dwarfs \citep{Lisker2006AJ....132.2432L, Lelli2022NatAs...6...35L}, suggesting that UGCA~337 might have been stripped of gas in the past.   

 %As a result, we can derive the total gas mass $\rm M_{gas}$ of $\rm (4.7\pm0.5)\times10^6\,M_\odot$ by multiplying by a factor of 1.33 after accounting for helium's contribution.
%in Figure~\ref{relocategalaxies}, UGCA~337 demonstrates a lower gas fraction, akin to the gas-poor dwarf galaxies discovered by \citet{Hu+etal+2023arXiv230905962H} utilizing FAST in the high-density environment. It indicates that there was gas stripping of UGCA~337 in the past \citep{Hu+etal+2023arXiv230905962H}. 

%\begin{figure}
%   \centering
%  \includegraphics[angle=0,width=0.7\columnwidth]{./MHI-Mstar.pdf}
%   \caption{ H\,{\sc i}-to-stellar mass ratio vs. galaxy stellar mass. The cyan dots denote the $\sim 30000$ ALFALFA-SDSS galaxies \citep{Durbala+etal+2020AJ....160..271D}, the yellow, magenta and blue hexagons represent the spiral, dwarf and irregular LSB galaxies \citep{Honey+etal+2018}, and the red square marks the galaxy UGCA~337 in this work labeled with its name. }
%   \label{relocategalaxies}
%\end{figure}

\section{Discussion} \label{sec:discussion}
%\subsection{Possibly identified a gas-poor LSB dwarf galaxy UGCA 337} \label{UGCA337}This stellar stream was also captured by the deep optical image of NGC~5055 in Figure~\ref{DeepOptical}. 
\subsection{Origin of the possible tidal gas}\label{subsec:extaplanar}
Our new high-sensitivity \HI\ mapping observations with FAST toward NGC~5055 reveal faint \HI\ clouds (clouds c1, 2, 3, and H1) at the edges of the \HI\ disk of NGC~5055. Clouds c2 and H1 were detected for the first time by us, and cloud c3 was identified for the first time as well. These four \HI\ clouds have a total \HI\ mass of $\rm \sim6.3\times10^7\, M_\odot$, contributing to less than one percent \HI\ mass of NGC~5055. In kinematics, clouds c1, 2, 3 look like the extensions of the spiral arms of NGC~5055 as shown in Figure~\ref{pv} and Figure~\ref{velfield}, resembling the tidal tails observed in other galaxies \citep[e.g.][]{Xu+etal+2021, Zhou+etal+2023}. In this section, we will mainly discuss the possible origins of them.

Tidal features generally indicate recent or past gravitational interactions/disruptions of galaxies \citep{Sancisi+etal+2008, Xu+etal+2021, Zhou+etal+2023}. However, UGC~8313 and UGCA~337 seem to lack the necessary mass to extract gas from their host galaxy NGC~5055, as indicated by the significant stellar mass ratios between NGC~5055 and UGC~8313 and UGCA~337 of about 274 and 306, respectively. Since NGC~5055 and UGC~8313 have a total stellar mass of $\rm 5.2\times10^{10}\,M_\odot$ and $\rm 1.9\pm0.6 \times10^8\,M_\odot$ at the distance of 8.87 Mpc respectively \citep{Carlsten+etal+2022}. In addition to UGC~8313 and UGCA~337, \citet{Chonis+etal+2011} proposed a minor interaction that NGC~5055 may have disrupted a missed dwarf galaxy in the last few Gyr, because they found a faint and huge arc-loop coherent stellar stream with a dim break at the north and northeast of NGC~5055. \citet{Chonis+etal+2011} roughly estimated a total mass of approximately $\rm 10^8\,M_\odot$ for this disrupted dwarf satellite. This mass appears also insufficient to extract gas from the host galaxy NGC~5055. Clouds 1, 2, and 3 appear to be unlikely outcomes resulting from the minor interaction between the disrupted dwarf galaxy/UGC 8313/UGCA 337 and NGC 5055. 

Alternatively, the tidal tails of NGC 5055 might be remnants of a past merger event that formed the galaxy NGC~5055. This is supported by the presence of the notable warped features, which may have resulted from a merger event \citep{Barnes2002MNRAS.333..481B}. In addition, it has been also discovered that the \HI\ kinematic features in NGC~5055 resemble those observed in M94 \citep{Zhou+etal+2023}. Specifically, a central, isolated host galaxy is surrounded by diffuse \HI\ tidal(s) and isolated \HI\ cloud(s). This similarity appears to suggest a common origin for the diffuse \HI\ features outside the host galaxy that the clouds could be part of the remnant of a major merger event in the past.

\section{Conclusions} \label{sec:conclusion}
We have performed a high-sensitivity \HI\ mapping observation toward the NGC~5055 galaxy group over an area of $1.^\circ5\times0.^\circ75$ with FAST.
We discovered a more extended gaseous disk of NGC~5055 compared to the \HI\ mappings of the WSRT. The FAST observation allows us to probe the \HI\ disk of NGC~5055 out to $ 23.'9$ ($\rm 61.7\,kpc$), and find a total \HI\ mass of $\rm\sim 1.1\times10^{10}\,M_\odot$. The latter is almost $22.4\%$ larger than that revealed by the pb-corrected HALOGAS survey, which is due to the fact that FAST is a single-dish telescope that dose not suffer from the zero-spacing problem, and then missing fluxes. We identified three \HI\ clouds with tidal features on the southeastern edge of the \HI\ disk of NGC~5055. They have a total \HI\ masses of $\rm \sim 2.7\times10^{7}\,M_\odot$, $\rm \sim 2.3\times10^{7}\,M_\odot$ and $\rm \sim 1.2\times10^{7}\,M_\odot$ respectively, and might be the remnants of a merger event in the past. A candidate high velocity cloud, H1, is confirmed to the north of NGC~5055 and near the minor axis. It has a total \HI\ mass of $\rm \sim 1.2\times10^6\,M_\odot$. In addition, the HI content of the dwarf galaxy UGCA~337 is robustly determined by the FAST observation for the first time, which has a narrow HI linewidth of $W_{50}=17.4\pm3.8$ km s$^{-1}$ with a total \HI\ mass of $\rm (3.5\pm0.3)\times10^6\,M_\odot$. Comparing the gas content and g-r color of UGCA~337 with typical low-mass dwarf galaxies, it appears relatively gas-poor despite its blue color, suggesting that UGCA~337 may have undergone gas stripping in the past. 

\begin{acknowledgements}
This work is supported by the National Key R\&D Program of China (2022YFA1602901) and the National Natural Science Foundation of China (Grants No. 12373001). This work was supported by the Open Project Program of the Key Laboratory of FAST, NAOC, Chinese Academy of Sciences.

\end{acknowledgements}

\bibliographystyle{raa}
\bibliography{reference0}

\label{lastpage}
\end{document}